# Anisotropic spin-relaxation induced by surface spin-orbit effects

Chao Zhou,[1,‡] Fatih Kandaz,[2,‡] Yunjiao Cai,[2] Chuan Qin,[2] Mengwen Jia,[1] Zhe Yuan,[3,*] Yizheng Wu,[1,4,*] and Yi Ji[2,*]

1. Department of Physics, State Key Laboratory of Surface Physics, Fudan University, Shanghai 200433, P.R. China

2. Department of Physics and Astronomy, University of Delaware, Newark, Delaware 19716, U.S.A.

3. The Center for Advanced Quantum Studies and Department of Physics, Beijing Normal University, Beijing 100875, P.R. China

4. Collaborative Innovation Center of Advanced Microstructures, Nanjing 210093, P.R. China

*Corresponding authors, zyuan@bnu.edu.cn; wuyizheng@fudan.edu.cn; yji@udel.edu.

‡These authors contributed equally.

**ABSTRACT**

It is a common perception that the transport of a spin current in polycrystalline metal is isotropic and independent of the polarization direction, even though spin current is a tensor-like quantity and its polarization direction is a key variable. We demonstrate surprising anisotropic spin-relaxation in mesoscopic polycrystalline Cu channels in nonlocal spin valves. For directions in the substrate plane, the spin-relaxation length is longer for spins parallel to the Cu channel than for spins perpendicular to it, by as much as 9% at 10 K. Spin-orbit effects on the surfaces of Cu channels can account for this anisotropic spin-relaxation. The finding suggests novel tunability of spin current, not only by its polarization direction but also by electrostatic gating.

## I. INTRODUCTION

Spin current, which is essential to spintronics technology, is a tensor-like quantity describing a flow of spin angular momenta with a polarization direction.[1] Tunability of spin current has been a desired functionality since the inception of spintronics,[2] but it remains a major challenge despite some promising progress.[3] Unlike an electrical current, a spin current decays as it propagates through a material because of the ubiquitous spin-relaxation. Spin-relaxation length, which characterizes the effective transport distance of a spin current, is a crucial quantity for describing many emergent phenomena, such as spin-Hall magnetoresistance,[4-6] spin Seebeck effect,[7] and spin pumping.[8] If spin-relaxation length depends on polarization direction, as the tensor-like nature of spin current would suggest, then the intriguing technological prospect arises that spin current could be tuned by polarization direction through the seemingly undesirable process of spin-relaxation.

According to theories, such anisotropic spin-relaxation could arise from various types of spin-orbit (SO) effects in semiconductors,[9, 10] graphene,[11, 12] or crystalline metals,[13] and is relevant to the fundamental question of whether the spin relaxation is of the Elliot-Yafet[14, 15] or Dyakonov-Perel type.[16] The anisotropic spin-relaxation time was observed experimentally in low-dimensional semiconductor systems[17-19] and attributed to the interplay of various SO contributions.[20-23] Using transport measurements in graphene-based non-local spin valves (NLSVs), Tombros *et al*.[24] claimed anisotropic spin signals in graphene for in-plane and out-of-plane polarization directions, but their claim was disputed and the result attributed instead to magnetoresistance effect in graphene with low carrier density under a strong out-of-plane magnetic field.[25] Therefore, anisotropic spin-relaxation in spin transport processes is still an open question and its unambiguous demonstration is desirable for the application of spin current in spintronic devices. Furthermore, polycrystalline metals such as Cu have not been considered for the study of anisotropic spin-relaxation, because of the lack of SO coupling and crystalline anisotropy.

In this work, anisotropic spin-relaxation is demonstrated in mesoscopic polycrystalline Cu channels using NLSV,[26-36] which is an ideal system for exploring spin-relaxation. The propagation of a pure spin current along a mesoscopic channel is well separated from the spin injection and detection processes. The exceptional signal-to-noise ratio of the nonlocal method allows a robust investigation of the spin-relaxation over a broad range of spin-transport distances. Anisotropic spin-relaxation in Cu is identified by exploring the dependence of the anisotropic signals on the spin-transport distance of the Cu channels, and it can be attributed to the SO effects of the Rashba-Sheka-Vasko type[21, 22, 37] on the surfaces of the Cu channels. The anisotropic differences in spin-relaxation lengths at 10 K are estimated to be as great as 9%. Because of the tunability of surface SO effect by an electric field perpendicular to surface,[38, 39] our finding introduces the attractive prospect of modulating spin currents via an electrostatic gate, as was originally proposed by Datta and Das for the pioneering spin transistors.[2]

## II. EXPERIMENTS

Figure 1(a) shows a scanning electron microscope (SEM) image of an NLSV that consists of a spin injector ($F_1$) and a spin detector ($F_2$) orientated in the $x$ direction and a Cu channel in the $y$ direction of the sample coordinate. $F_1$ and $F_2$ are made of Permalloy (Py), an alloy of NiFe, and have widths of ~150 nm and ~130 nm and thicknesses of 18 nm and 12 nm, respectively. The widths of the Cu channel are between 160 and 210 nm, and the center-to-center distance $L$ between $F_1$ and $F_2$ ranges from 400 to 1050 nm. Two thicknesses of Cu, 200 nm and 110 nm, are used. A layer of 3 nm $AlO_x$ is placed at the interfaces between Py and Cu for efficient spin injection and detection.[40] The precise geometries of all the NLSVs are measured by SEM after transport measurements for data analysis.

A charge current $I_e$ between $F_1$ (I+) and the upper end of the Cu channel (I−) injects a spin accumulation into the channel. The gradient of the spin accumulation drives a pure spin current along the channel in the −$y$ direction. A nonlocal voltage $V_{nl}$ is measured between $F_2$ (V+) and the lower end of the Cu channel (V−). The nonlocal signal $R_s = V_{nl}/I_e$ versus $B_x$, a magnetic field applied in the $x$ direction, is shown in Figure 1(b) for an NLSV. The $R_s$ alternates between a high-value state, corresponding to parallel (P) spins of $F_1$ and $F_2$, and a low-value antiparallel (AP) state, yielding a spin signal of $\Delta R_s^x = 6.4$ mΩ for the $x$-spins. The average $R_s$ of the P and AP states is the "baseline" and corresponds to a null spin accumulation in the Cu channel. The P state $R_s$ is higher than the baseline by $\Delta R_s^x/2$, and the AP state $R_s$ is lower by the same amount.

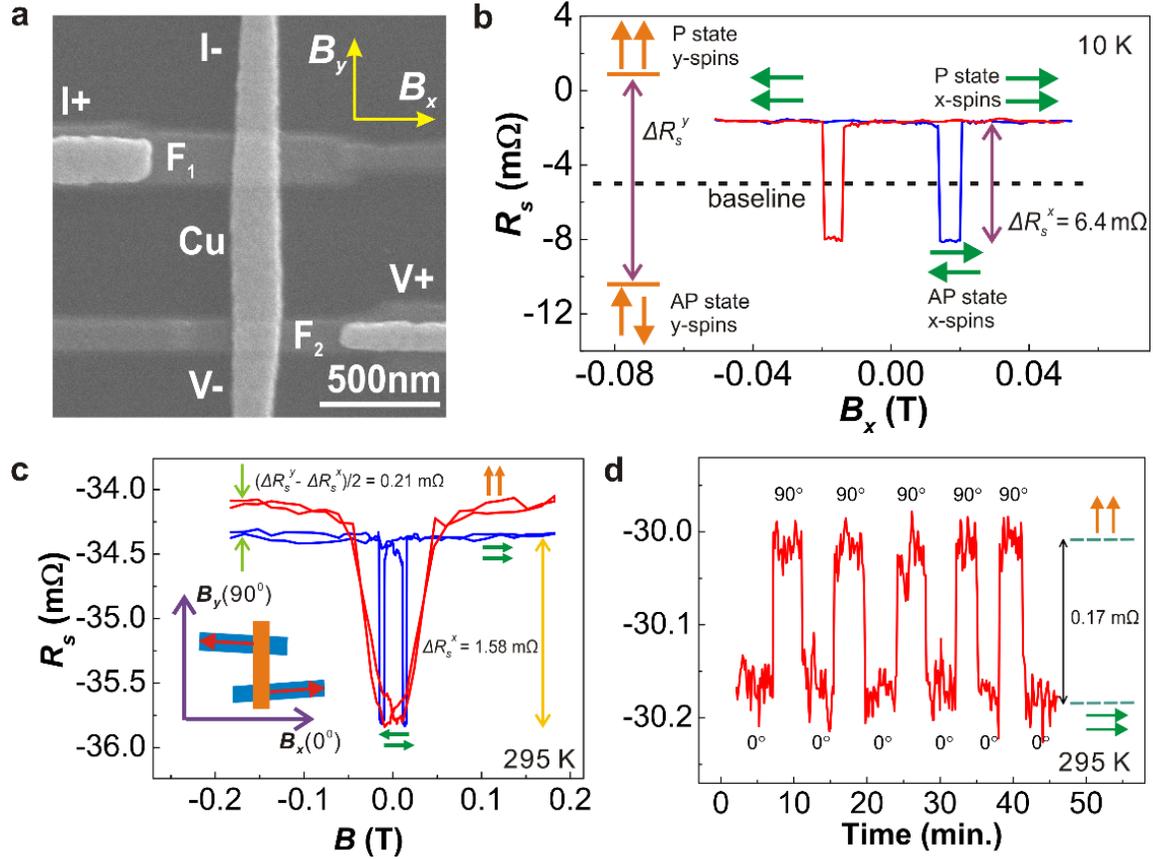

**Figure 1.** (a) SEM image an NLSV and the measurement configuration. (b) The $R_s$ versus $B_x$ curve for a typical NLSV at 10 K, and illustration of P and AP states for $x$-spins and $y$-spins. The $R_s$ values of the P state and AP state are symmetrical to the baseline. (c) The $R_s$ versus $B$ curves for another NLSV at 295K, with field applied along the $x$ direction (blue) and the $y$ direction (red). The inset shows that F$_1$ and F$_2$ of this NLSV are designed to make small but opposite angles (±3°) with the $x$ direction. (d) The $R_s$ versus time plot at 295 K while the NLSV [the same as in (c)] is rotated between 0° and 90° under a 0.2 T field.

When the magnetic field is applied in the $y$ direction to polarize the spins of F$_1$ and F$_2$, in principle the spin signals for $y$-spins can also be detected. The P and AP states for the $y$-spins, if both can be reached, should also be symmetrical about the baseline, and the difference between two states is the spin signal $\Delta R_s^y$ for $y$-spins, as illustrated in Figure 1(b). If $\Delta R_s^y \neq \Delta R_s^x$, the spin signals are anisotropic and we can define the anisotropic signal $\left(\Delta R_s^y - \Delta R_s^x\right)$ and the percentage anisotropic signal $(\Delta R_S^y - \Delta R_S^x)/\Delta R_S^x$. When $\Delta R_S^y > \Delta R_S^x$, the P state of the $y$-spins is higher than that of the $x$-spins by $(\Delta R_S^y - \Delta R_S^x)/2$. In this work, we measure this quantity by directly comparing the P states of $x$-spins and $y$-spins. Note that it is often not a straightforward matter to access the AP state of $y$-spins, because the easy axis of $F_1$ and $F_2$ lies in the $x$ direction. The baselines for $x$-spins and $y$-spins should be identical for the same NLSV, at a fixed temperature, and wired in the same circuit.

The first evidence of anisotropic signals is obtained at 295 K with a specially designed NLSV, in which $F_1$ and $F_2$ have opposite small angles (±3°) with respect to the $+x$ axis, as shown in the inset of Figure 1(c). Figure 1(c) shows the measured $R_s$ as a function of the applied magnetic fields $B_y$ in the $y$ direction (red curve) and $B_x$ in the $x$ direction (blue curve). Obviously the red curve shows greater overall $R_s$ variation than the $x$-spin signal $\Delta R_S^x = 1.58$ mΩ of the blue curve. The spins of $F_1$ and $F_2$ are aligned to the P state in the $+y$ direction when $B_y = +0.2$ T. As the field $B_y$ is reduced, the spins gradually rotate toward the easy axis (±$x$ direction). Because of the opposite angles, the $F_1$ and $F_2$ spins snap into the $-x$ and $+x$ directions respectively when $B_y = 0$, reaching an $x$-spin AP state. The lowest values of both the red and blue curves correspond to the same AP state for $x$-spins. The highest values of the two curves correspond to the P states for the $y$- and $x$-spins, and the difference gives $(\Delta R_S^y - \Delta R_S^x)/2 = 0.21$ mΩ. The percentage anisotropic signal is $(\Delta R_S^y - \Delta R_S^x)/\Delta R_S^x = 26\%$.

It is obvious from Figure 1(c) that a field of 0.2 T is sufficient to align the spins into the P state along the hard axis (±$y$ direction). To further confirm the anisotropic signal, we rotate the same NLSV in a 0.2 T magnetic field between 0° and 90° periodically to alternate the field, and thereby the aligned spins, between the $x$ and $y$ directions of the sample coordinate, respectively.

A periodic $R_s$ change of 0.17 mΩ is observed, as shown in Figure 1(d), with the P state of the $y$-spins (90°) higher than the P state of $x$-spins (0°), confirming the anisotropic signal. A small baseline change of −0.03 mΩ, induced by subtle changes of AC coupling in the measurement circuits while the sample rotates, has to be subtracted to obtain an accurate anisotropic signal. The calibrated value, $(\Delta R_s^y - \Delta R_s^x)/2 = [0.17 - (-0.03)]$ mΩ $= 0.20$ mΩ, is consistent with the 0.21 mΩ shown in Figure 1(c). More details on the baseline changes are given in Notes S1 of the Supplemental Material.[41]

Systematic measurements of anisotropic signals for NLSVs of various channel lengths $L$ are carried out in a probe-station at 10 K. These NLSVs are fabricated on two substrates (samples), A and B, with Cu thicknesses of 200 nm and 110 nm, respectively. Both substrates are Si covered with 300 nm SiN. NLSVs on the same substrate undergo identical processing procedures. We alternate the magnetic field between the $x$ and $y$ directions to access the P states of the $x$-spins and $y$-spins, respectively. The rotating magnetic field is realized by a bias field $B_{bias} \sim 0.03$ T in the $x$ direction from a permanent magnet and a variable applied field $B_y$ in the $y$ direction from an electromagnet, as shown in Figure 2(a). The purpose of the $B_{bias}$ is to align $F_1$ and $F_2$ into a good $x$-spin P state under zero applied field. To verify this, the field $B_x$ is applied in the same direction (+$x$) as the $B_{bias}$, and the measured $R_s$ versus $B_x$ curve on a typical NLSV is shown in Figure 2(b). The bias field shifts the curve along the $B_x$ axis, and the $R_s$ value at $B_x = 0$ T clearly indicates a good $x$-spin P state. The difference of $\Delta R_s$ (the magnitudes of the dips) between the red and blue curves is attributed to the proximity of the $F_1$ and $F_2$ switching fields and the resultant mediocre AP state while the field ramps downward. The signal of the upward branch is taken as the full spin signal for $x$-spins: $\Delta R_s^x = 10.7$ mΩ.

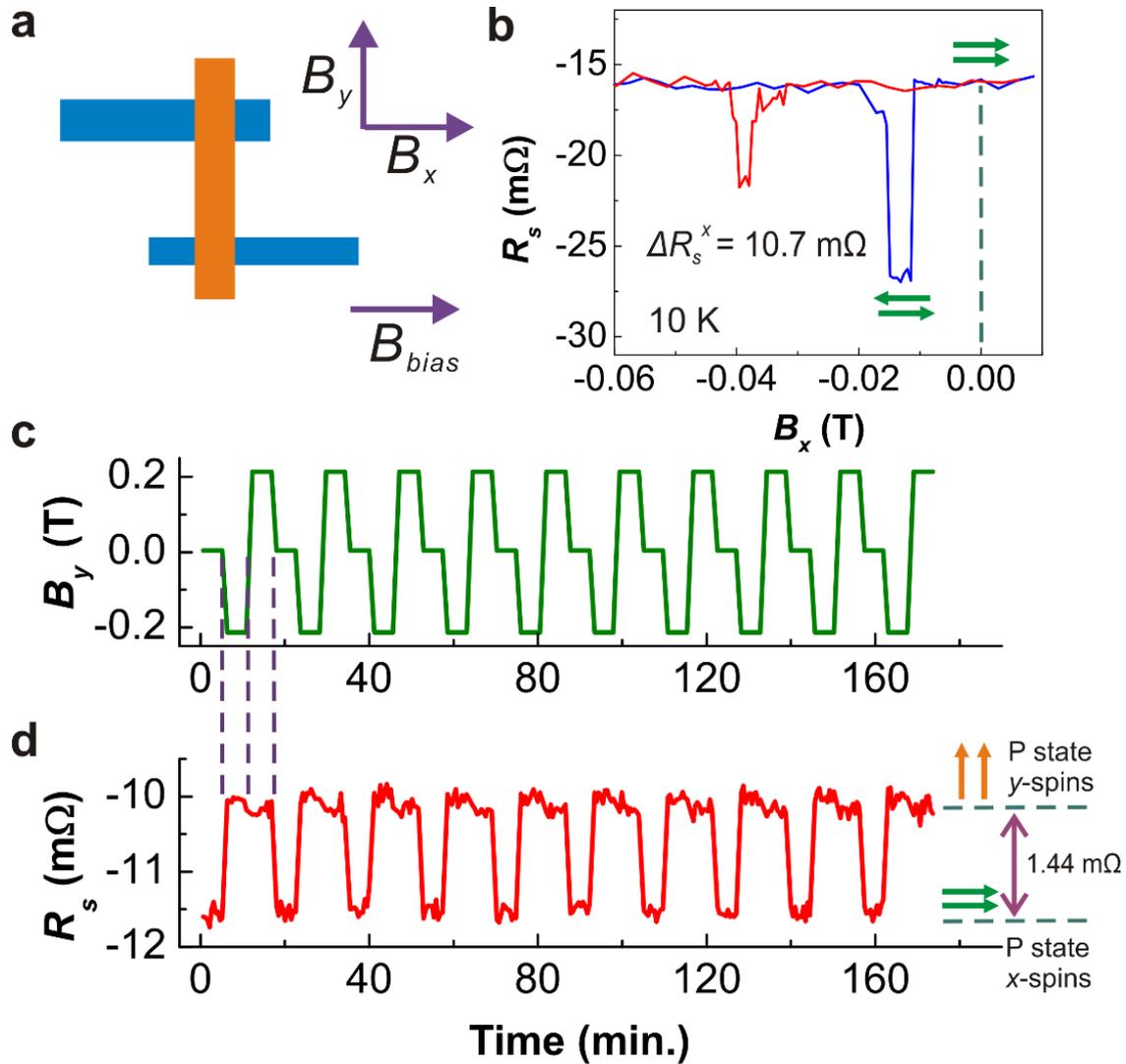

**Figure 2.** (a) Illustration of a fixed bias field $B_{bias}$ in the $x$ direction and an applied field that can be orientated either in the $x$ direction or in the $y$ direction. (b) The $R_s$ versus $B_x$ curve with applied field in the $x$ direction in the presence of the bias field. (c) The applied field $B_y$ in $y$ direction and (d) $R_s$ as a function of time.

Figure 2(c) illustrates that $B_y$ alternates periodically between 0 T, −0.2 T, and +0.2 T while $B_{bias}$ remains in the $x$ direction. As a result, the aligned spins alternate between the +$x$, −$y$, and +$y$ directions, respectively. Because the applied fields $B_y = \pm 0.2$ T are much larger than the 0.03 T bias field, they are sufficient to align spins into the ±$y$ directions. Figure 2(d) illustrates the measured $R_s$ as a function of time. When $B_y$ changes from 0 to −0.2 T, $R_s$ increases abruptly by ~1.5 mΩ. When $B_y$ changes from −0.2 T to +0.2 T, $R_s$ stays at almost the same value. When $B_y$ changes from +0.2 T to 0 T, $R_s$ decreases by ~1.5 mΩ. As the field cycles go on, the same pattern is consistently reproduced and again provides clear evidence for an anisotropic signal. We take the average $R_s$ values at ±0.2 T as the P state of the y-spins and the $R_s$ value at 0 T as the P state of the x-spins. We extract the anisotropic signal, $(\Delta R_s^y - \Delta R_s^x)/2 = 1.44$ mΩ, by averaging the difference between two states over many cycles. The percentage anisotropic signal is $(\Delta R_s^y - \Delta R_s^x)/\Delta R_s^x = 26.8\%$.

The quantitative description of the spin signals is useful for identifying the origin of the anisotropy. The spin signal of NLSVs with oxide interfaces is well described by[28, 30, 40, 42]

$$\Delta R_s = \frac{P_1 P_2 \rho \lambda}{A} e^{-\frac{L}{\lambda}} \quad (1),$$

where $P_1$ and $P_2$ are the effective spin polarizations of F$_1$ and F$_2$, $\rho$ is the resistivity of Cu, $\lambda$ is the Cu spin-relaxation length, and $A$ is the cross-sectional area of Cu. In NLSV studies,[30, 34-36] it is customary to assume that $P_1 = P_2 = P$ when F$_1$ and F$_2$ are made of same material (Py) and have similar dimensions. Because $A$ and $\rho$ are spin-independent quantities, the observed anisotropic spin signals should originate from either $P$ or $\lambda$. F$_1$ and F$_2$ are polycrystalline, and the injection current is nearly perpendicular to the interface. Therefore, there is no reason for an anisotropic $P$ between the x-spins and y-spins, and our results suggest an anisotropic $\lambda$. Moreover, if the $P_x$ for x-spins and $P_y$ for y-spins differ by a fixed percentage $(P_y - P_x)/P_x$, the percentage anisotropic signal $(\Delta R_s^y - \Delta R_s^x)/\Delta R_s^x$ should remain a constant and be independent of channel length $L$, according to Eq. (1). However, if the $\lambda_x$ for x-spins and the $\lambda_y$ for y-spins differ by a fixed percentage $(\lambda_y - \lambda_x)/\lambda_x$ due to the anisotropic spin-relaxation, $(\Delta R_s^y - \Delta R_s^x)/\Delta R_s^x$ should increase as a function of $L$.

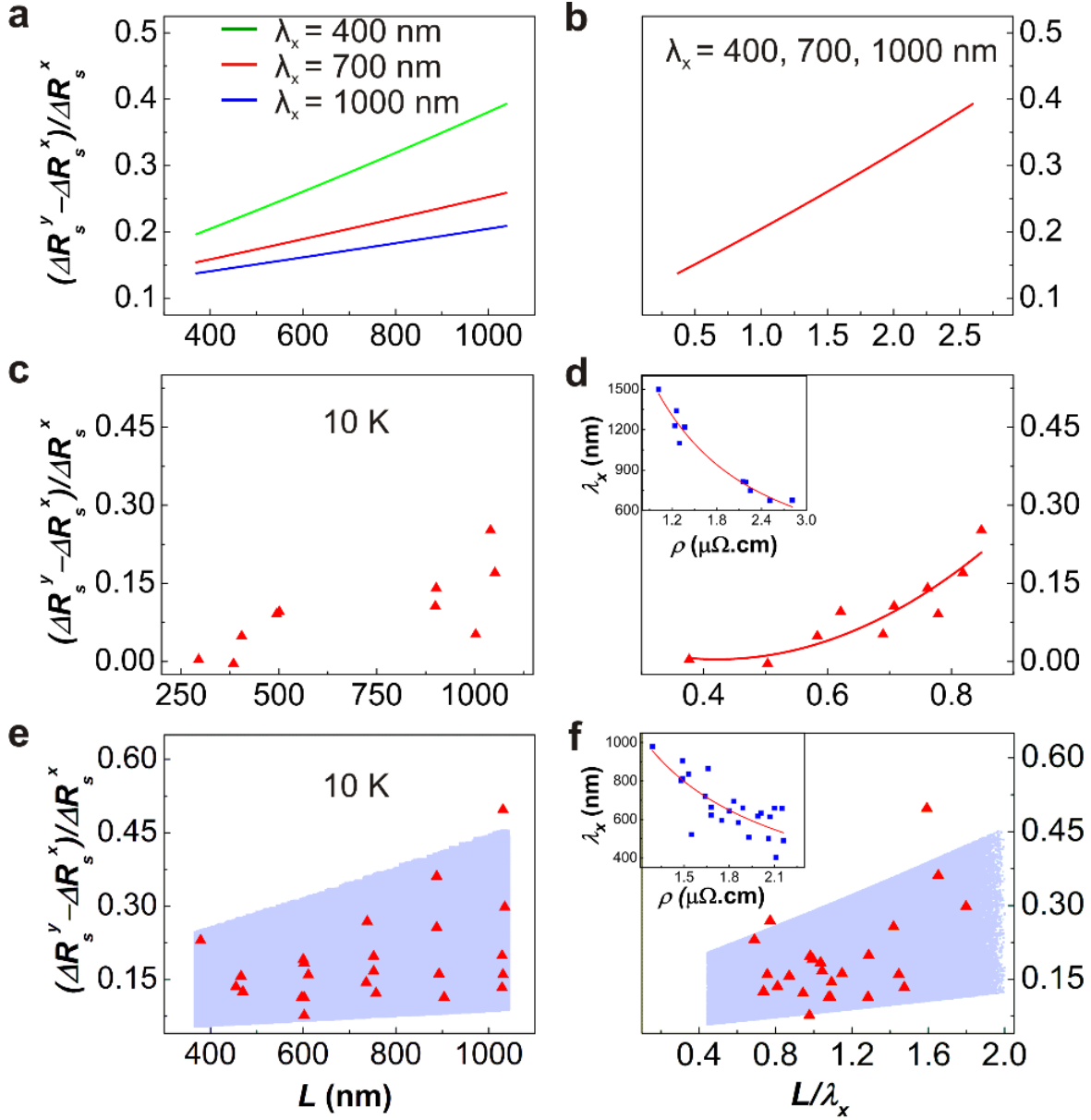

**Figure 3.** (a) Calculated $(\Delta R_s^y - \Delta R_s^x)/\Delta R_s^x$ as a function of $L$ for various values of $\lambda_x$. It is assumed that $(\lambda_y - \lambda_x)/\lambda_x$ = 10%. (b) The same results as in (a) are plotted as a function of $L/\lambda_x$. Three curves collapse onto one. $(\Delta R_s^y - \Delta R_s^x)/\Delta R_s^x$ versus (c) $L$ and (d) $L/\lambda_x$ at 10 K for NLSVs on substrate A. The solid line in (d) is a guide to the eyes. $(\Delta R_s^y - \Delta R_s^x)/\Delta R_s^x$ versus (e) $L$ and (f) $L/\lambda_x$ at 10 K for NLSVs on substrate B. The insets in (d) and (f) show the $\lambda_x$ versus $\rho$ at 10 K and the fits (solid lines) by the Elliott-Yafet model for substrates A and B, respectively. The shaded areas in (e) and (f) are calculation by randomized parameters in the range: $\lambda_x$ = 670 ± 180 nm and $(\lambda_y - \lambda_x)/\lambda_x$ = (9±5) %.

Figure 3(a) shows the calculated $(\Delta R_s^y - \Delta R_s^x)/\Delta R_s^x$ as a function of $L$ from Eq. (1), assuming $(\lambda_y - \lambda_x)/\lambda_x$ = 10% for several $\lambda_x$ values. The calculated $(\Delta R_s^y - \Delta R_s^x)/\Delta R_s^x$ increases monotonically with $L$. In Figure 3(b), a normalized channel length $L/\lambda_x$ is used as the horizontal axis. All three curves in Figure 3(a) collapse to a single curve in Figure 3(b), giving a more universal relationship that is immune to variation in $\lambda_x$. The experimental values of $(\Delta R_s^y - \Delta R_s^x)/\Delta R_s^x$ for NLSVs on substrates A and B are plotted as a function of $L$ in Figures 3(c) and (e), respectively. The highest value of $(\Delta R_s^y - \Delta R_s^x)/\Delta R_s^x$ at a particular $L$ increases as a function $L$ for both substrates. For substrate B, the highest percentage reaches ~46% for $L$ = 1050 nm.

Note that the simple monotonic increase predicted in Figures 3(a) and (b) is not obvious because of data dispersion, which can be attributed to variations in $\lambda$ between NLSVs in spite of their being fabricated under identical conditions. The $\lambda_x$ versus $\rho$ graphs for Cu are plotted in the insets of Figures 3(d) and 3(f) for NLSVs on substrates A and B, respectively. We measure $\rho$ for each NLSV by sending a current through the Cu and measuring a voltage between $F_1$ and $F_2$. We calculate $\lambda_x$ from the $\Delta R_s^x$ and $\rho$ of this NLSV using Eq. (1) and assuming $P$ = 20%, which is justified by previous measurements.[40, 43, 44] The data are fitted by $\lambda_x = \beta/\rho$, which is implied by the Elliot-Yafet model.[14, 15, 40] Here $\beta$ is the fitting parameter and is related to the spin-relaxation rate in the Cu. The fitted $\lambda_x$ corresponding to the measured $\rho$ is used to calculate the $L/\lambda_x$ for each NLSV. The $(\Delta R_s^y - \Delta R_s^x)/\Delta R_s^x$ is plotted as a function of $L/\lambda_x$ for the substrates A and B in Figure 3(d) and (f), respectively, and the increasing trends are clear. Furthermore, the data for A and B can be combined to generate a single plot, shown in Figure S2 in the Supplemental Material,[41] yielding a clear increase of $(\Delta R_s^y - \Delta R_s^x)/\Delta R_s^x$ versus $L/\lambda_x$ that is qualitatively consistent with the calculation in Figure 3(a) and (b). Therefore, we conclude that the spin-relaxation is anisotropic in the Cu channels and causes the observed anisotropic spin signals.

In Figure 3(f), there is still substantial data dispersion, which can be explained by variations in $(\lambda_y - \lambda_x)/\lambda_x$ in addition to the variations in $\lambda_x$. Using randomized parameters in the range of $\lambda_x = 670 \pm 180$ nm and $(\lambda_y - \lambda_x)/\lambda_x = (9 \pm 5)\%$ for substrate B, we calculate the $(\Delta R_s^y - \Delta R_s^x)/\Delta R_s^x$ versus $L$ and $L/\lambda_x$. These are shown as the shaded regions in Figures 3(e) and 3(f), respectively. The dispersion of the experimental data is qualitatively reproduced. The $\lambda_y$ can be estimated from the $\Delta R_s^y$ of each device. The values of $\lambda_y - \lambda_x$ and $(\lambda_y - \lambda_x)/\lambda_x$ are plotted as functions of the $\lambda_x$ value for NLSVs on A and B in Figure S3 of the Supplemental Material.[41] The average values are $(\lambda_y - \lambda_x)/\lambda_x = (5.3 \pm 4.1)\%$ for substrate A and $(\lambda_y - \lambda_x)/\lambda_x = (8.6 \pm 3.3)\%$ for substrate B.

The magnetoresistance effect that can mimic anisotropic spin signals in graphene[24, 25] can be ruled out in this study. We use a small field ($\leq 0.2$ T) to rotate spins in the substrate plane. In Figure 1(c), the anisotropic signal difference between the red and blue curves is already quite obvious even at 0.05 T, and saturates at higher fields. Furthermore, the metallic Cu channel has high carrier density and is not easily influenced by the applied magnetic field.

### III. DISCUSSION

All the relevant theories link anisotropic spin-relaxation to various SO couplings,[9-13] which should account for the observed anisotropic-spin relaxation in Cu channels as well. Furthermore, we can rule out the contribution of bulk SO to the observed anisotropic spin relaxation by performing *ab initio* transport calculation of spin-relaxation in bulk Cu. We consider a Py(10nm)/Cu(100nm) bilayer and an electric current perpendicular to the Py/Cu interface for spin injection. The spin accumulation in the Cu is plotted in Figure 4(a) as a function of the distance to the Py/Cu interface. It exhibits an exponential decay following the spin diffusion theory.[45] With the real SO strength of bulk Cu, $\lambda$ is found to be larger than 400 nm. If we artificially increase the SO strength of Cu by a factor of seven, the calculated $\lambda$ is reduced to ~16 nm. In both cases, however, the spin accumulation at various positions is independent of the polarization directions, indicating that the bulk SO cannot account for the anisotropic spin-relaxation in the Cu channels. More details of the calculation are given in Notes S4 in the Supplemental Material.[41]

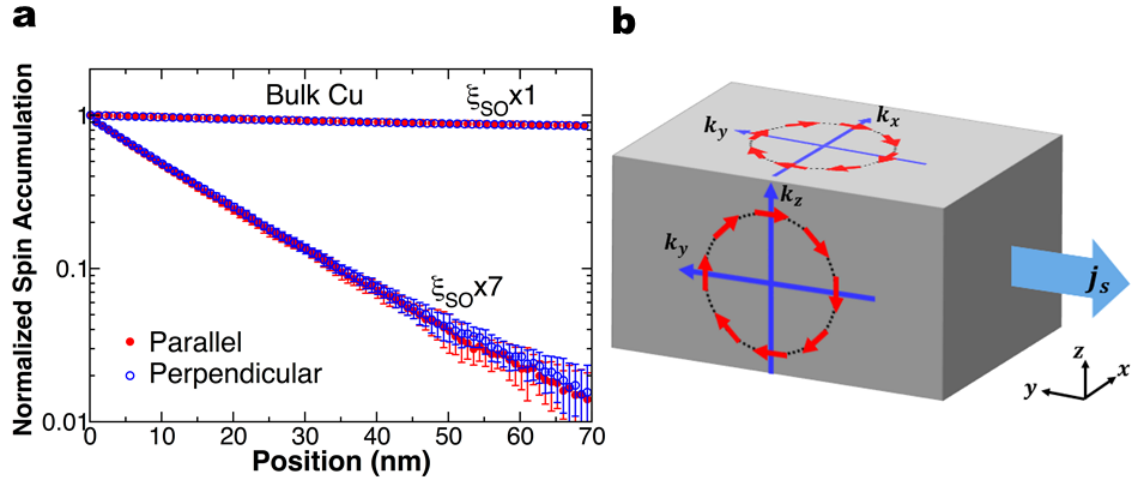

**Figure 4.** (a) Calculated spin accumulation in bulk Cu as a function of distance from the Py/Cu interface. While the upper data are obtained using the real SO strength of bulk Cu, the lower ones are calculated with the SO strength artificially increased by a factor of seven. In both cases, the spins parallel to the spin-current direction ($y$-spins, red symbols) are the same as the spins perpendicular to it ($x$-spins, blue symbols) indicating the bulk SO interaction does not introduce the anisotropic spin-relaxation. (b) Sketch of the SO effective magnetic fields at the side and top surfaces of a Cu channel.

Besides bulk, SO effects could also be present at surfaces and interfaces. Our experiments were carried out at low temperatures, where phonon scattering is mostly suppressed. The electrical resistivity arises mainly from defect and surface scattering. The surface SO effect,[37] originating from inversion asymmetry across a surface, results in spin-relaxation when conduction electrons are scattered from the surfaces of the mesoscopic Cu channels. For two-dimensional graphene,[24, 25] the SO effect on the top and bottom surfaces can induce anisotropy between out-of-plane polarization (z-spins) and in-plane polarization (x-spins and y-spins), but will not induce anisotropy between x-spins and y-spins. However, the Cu channels in our experiments have comparable width and thickness. Therefore, the SO contribution from the side surfaces should be comparable to that from the top and bottom surfaces, and could account for the observed anisotropy between x-spins and y-spins.

The SO Hamiltonian of the side surfaces lying in the y-z plane can be expressed as:

$$H_R^S = \frac{\alpha_R}{\hbar} \boldsymbol{\sigma} \cdot (\hat{x} \times \boldsymbol{p}) = \frac{\alpha_R}{\hbar} (p_y \sigma_z - p_z \sigma_y) = \frac{\alpha_R}{\hbar} \begin{pmatrix} p_y & ip_z \\ -ip_z & -p_y \end{pmatrix}, \quad (2)$$

where the $\alpha_R$ is the surface SO strength, $\boldsymbol{\sigma}$ is the Pauli matrix and $\boldsymbol{p}$ is the electron momentum. This type of SO interaction was derived by Rashba and Sheka[21] for wurtzite type crystals, and then by Vasko[22] for two dimensional electron gas. We define the eigenstates of $\sigma_x$ and $\sigma_y$, i.e. $|\pm x\rangle = (1, \pm 1)/\sqrt{2}$ and $|\pm y\rangle = (1, \pm i)/\sqrt{2}$, as $\pm x$-spins and $\pm y$-spins, respectively. The rates of spin-relaxation are inversely proportional to the spin relaxation time $\tau_{x(y)}$ and can be calculated perturbatively from the SO Hamiltonian (Supplemental Material, Notes S5).[41] Specifically, we have $\frac{1}{\tau_x} \propto \int d^3p |\langle -x|H_R|+x\rangle|^2 = \int d^3p \cdot (p_y^2 + p_z^2)$ and $\frac{1}{\tau_y} \propto \int d^3p |\langle -y|H_R|+y\rangle|^2 = \int d^3p \cdot p_y^2$. The x-spins have a higher spin-relaxation rate than the y-spins in the presence of the SO Hamiltonian (Eq. (2)) in agreement with the experiments. The influence of the SO effect at the top and bottom surfaces lying in the x-y plane can be evaluated in the same manner, and it makes the same contribution to the spin-relaxation rates for x-spins and y-spins (Supplemental Material, Notes S5).[41] Therefore, the observed anisotropic spin-relaxation is a result of the SO interaction at the side surfaces of the Cu channels. Large SO effects on Cu (111) surfaces were experimentally observed[46] and theoretically justified.[47, 48] For the polycrystalline NLSVs, high rates of spin relaxation on the surfaces of Cu channels[49] and SO effects on the surfaces of Ag channels covered with $Bi_2O_3$[50] were reported.

The SO effective magnetic field is parallel to the surface and depends on the direction of electron momentum, as sketched in Figure 4(b). When an electron is moving in the vicinity of a surface, the effective magnetic field tends to align the electron spin with the direction of the field, and therefore causes relaxation of spins that are not aligned with the field. For electrons near the side surfaces, the *x*-spins are always perpendicular to the SO effective magnetic field of the side surfaces and consequently have a higher probability of spin-relaxation than the *y*-spins, which have finite components in the direction of the effective magnetic fields. For the top and bottom surfaces, the SO effective magnetic field rotates through the *x-y* plane and has the same effect on the spin relaxation for *y*-spins and *x*-spins. The SO effective electric field is perpendicular to the surfaces. It was demonstrated that the strength of SO interaction can be tuned by an applied electric field perpendicular to the surface via an electrostatic gate,[38, 39] indicating the possibility of manipulating the spin-relaxation electrically.

The above interpretation does not depend on whether the spin-relaxation is of Elliott-Yafet[14, 15] (EY) or Dyakonov-Perel (DP) type.[16] The only assumption is that the electrons are under the influence of the SO effective magnetic field near the side surfaces. The spin-flip may occur upon momentum scattering (*i.e.* EY mechanism) or between momentum scattering events (*i.e.* DP mechanism). Both types would result in higher rates of spin-relaxation for *x*-spins than for *y*-spins when electrons moving near side surfaces. In principle, the EY and DP could coexist and contribute constructively to the anisotropic spin-relaxation. The mechanisms could also be different between transport in the bulk and transport near surfaces. While EY is commonly viewed as the dominant mechanism in metals with bulk inversion symmetry, recent experimental[51] and theoretical[52] works point to the possibility of DP mechanism in Pt. Nevertheless, the $\lambda_x$ versus $\rho$ plots in the insets of Figure 3 (d) and (f) for our two samples can be fitted reasonably well by the EY model.

The electron mean free paths in the Cu channels are relevant in the discussion of surface and bulk scattering. The momentum relaxation time $\tau_e$ can be estimated from Cu resistivity $\rho$ using Drude model $\tau_e = m/(\rho n e^2)$, where $n = 8.47 \times 10^{28}$ m$^{-3}$ is Cu electron density,[53] $m$ is electron mass, and $e$ is electron charge. The mean free path $l$ is estimated by $l = v_F \tau_e$, where $v_F = 1.57 \times 10^6$ m/s is the Cu Fermi velocity.[53] The average $\rho$ values at 10 K for NLSVs on samples A and B are both 1.8 µΩ·cm, yielding $\tau_e = 23.3$ fs and $l = 37$ nm, which is obviously smaller than the Cu widths ($w = 175$ nm for A and $w = 207$ nm for B on average) and thicknesses (200 nm for A and 110 nm for B). Therefore, bulk momentum scattering dominates over surface scattering, and the electron transport is mainly diffusive. However, as the electrons diffuse in the Cu channel over a spin-relaxation length that is greater than the widths and thicknesses, there are still substantial surface scattering events.

The relation between the diffusion time $\tau_d$ and distance $d$ is given by $d^2 = D\tau_d$, where $D$ is the bulk diffusion constant. An application of this is $\lambda_{x(y)}^2 = D\tau_{x(y)}$, which links spin diffusion (relaxation) length and time. The diffusion constant can be evaluated by $D = 1/(\rho e^2 N)$ and $N = 3n/2E_F$, where $N$ is the Fermi level density of states and $E_F = 7.00$ eV is the Cu Fermi energy.[53] Electrons at the center of cross-section of Cu channel need to diffuse over $d = w/2$ in the $\pm x$ directions to reach the side surfaces. Using $d = w/2 = 88$ nm for sample A, the time for a diffusion distance of 88 nm in any direction is 405 fs. Considering the three possible dimensions for diffusion, the average time for electrons to diffuse 88 nm in $\pm x$ directions has to be increased by a factor of 3 yielding $\tau'_e = 1.22$ ps. The ratio $\tau'_e/\tau_e = 52$ indicates that an electron reaches a side surface after 52 bulk scattering events on average. From average spin diffusion length $\lambda_x$ = 1010 nm of sample A, we estimate the spin-relaxation time $\tau_x = 53.4$ ps, which is the average diffusion time for an electron before a spin-flip event. During time interval $\tau_x$, there are $\tau_x/\tau'_e = 44$ scattering events with the side surfaces. Similarly, we calculate ~34 scattering events with top and bottom surfaces of Cu channels (200 nm thick for A) during $\tau_x$. The number of bulk scattering events during $\tau_x$ is $\tau_x/\tau_e = 2292$. Therefore, the overall bulk-to-surface scattering ratio is $2292/(44 + 34) = 29$. The spin relaxation and its anisotropy can be accounted for by assuming bulk spin-flip probabilities in the order of $10^{-4}$ and surface spin-flip probabilities in the order of $10^{-3}$ to $10^{-2}$,

It is tempting to compare the magnitudes of the anisotropic effects between A and B, which have different thickness and thereby different areas of side surfaces. However, for a mainly diffusive Cu channel, a moderately reduced thickness actually forces the electrons to diffuse toward the side surfaces. Therefore, the anisotropic effect is not necessarily reduced for a thickness of 110 nm as compared to 200 nm. Quantitative comparison is challenging because of the various types of scattering and the variations of $\lambda$ and $\rho$ between devices. In Figure S2 of the Supplemental Material,[41] it is reassuring to our interpretation that all data points from two samples scale with $L/\lambda_x$. If the Cu channel thickness is more drastically reduced, *e.g.* to 10 nm, the scattering from top and bottom surfaces would be dominant and result in short spin diffusion length (< 100 nm), which makes NLSV measurements difficult. We propose for future work that many NLSVs can be fabricated under identical conditions with Cu channels that have same thickness but different widths. Furthermore, values of $\rho$ and $\lambda$ have to be carefully evaluated and taken into account in the analysis.

## IV. CONCLUSIONS

In conclusion, we have demonstrated anisotropic spin-relaxation in the mesoscopic metallic (Cu) channels of nonlocal spin valves. The anisotropic differences in spin-relaxation lengths are as great as 9% at 10 K. Consequently, the spin current in the Cu channel can be tuned by its polarization-direction. Surface spin-orbit effects of the Rashba-Sheka-Vasko type account for the observed anisotropy and offer the prospect of electrical tuning of spin currents via electrostatic gating.

## ACKNOWLEDGMENTS

Kandaz acknowledges support from the Republic of Turkey Ministry of National Education. Zhou, Jia, and Wu acknowledge support from the National Key Basic Research Program of China (Grants No. 2016YFA0300700, and No. 2015CB921401) and the National Science Foundation of China (Grants No. 11434003, No. 11474066, and No. 11734006). Yuan acknowledges support from National Science Foundation of China (Grant No. 61774018)

# Supplementary materials for

# Anisotropic spin-relaxation induced by surface spin-orbit effects

Chao Zhou,[1‡] Fatih Kandaz,[2‡] Yunjiao Cai,[2] Chuan Qin,[2] Mengwen Jia,[1] Zhe Yuan,[3*] Yizheng Wu,[1,4*] and Yi Ji[2*]

1. Department of Physics, State Key Laboratory of Surface Physics, Fudan University, Shanghai 200433, P.R. China

2. Department of Physics and Astronomy, University of Delaware, Newark, Delaware 19716, U.S.A.

3. The Center for Advanced Quantum Studies and Department of Physics, Beijing Normal University, Beijing 100875, P.R. China

4. Collaborative Innovation Center of Advanced Microstructures, Nanjing 210093, P.R. China

*Corresponding authors, zyuan@bnu.edu.cn; wuyizheng@fudan.edu.cn; yji@udel.edu.

‡These authors contributed equally.

# Supplementary Materials

**Notes S1. Baseline of the $R_s$ versus $B$ curves and implications on measurements.**

Figure 1(b) in the main article illustrates that the baseline is the average $R_s$ value for the parallel (P) and antiparallel (AP) states. The baseline arises from a background charge voltage that corresponds to a null spin accumulation in the NLSV. For strictly one-dimensional NLSVs, the baseline value should simply be zero because there is no charge voltage between the $F_2$ and the Cu channel. However, for realistic three-dimensional NLSV devices, the baseline $R_s$ value is not exactly zero because of the nontrivial current distribution in the mesoscopic structure.[53,54]

In addition, the nonlocal measurements are carried out using alternating current (AC) lock-in method, and any variation in AC coupling affects the baseline. For example, a change of AC frequency would alter the effective reactance (capacitance and inductance) values of the measurement circuits and thereby alter the baseline. Measurements carried out in separate circuits for the same NLSV would have different baselines, because the effective reactance values are different between circuits. Similarly, subtle changes in wiring configurations of the measurement circuits could result in variations of baseline values as well.

In the main article, the anisotropic signal $(\Delta R_s^y - \Delta R_s^x)/2$ is measured by detecting the difference between the P states for $y$-spins and $x$-spins, as illustrated in Figure 1 and 2. The underlying assumption is that the baseline for the NLSV remains unchanged when the spin states are altered between $x$ and $y$ directions. For an accurate detection of $(\Delta R_s^y - \Delta R_s^x)/2$, one has to either ensure a truly intact baseline during measurement or compensate for any change of baseline.

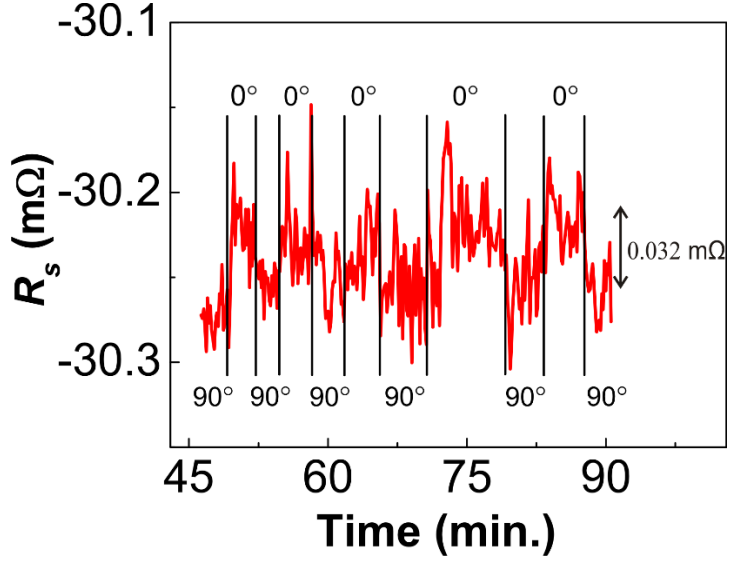

**Figure S1.** The plot of $R_s$ versus time for a NLSV, which is rotated between 0° and 90° back and forth under zero magnetic field. This measurement was carried out for the same NLSV and in the same circuit as those of Figure 1(d) in the main article.

The two $R_s$ versus $B$ curves in Figure 1(c) are measured in separate circuits and therefore the baseline values are not guaranteed to be the same. However, the lowest $R_s$ values for both curves are associated with AP state for $x$-spins and it can be used as a common reference point for comparison, as explained in the main article. We shift the two curves on the vertical scale so that the lowest points of both curves are aligned at the same $R_s$ value. Then the anisotropic signal $(\Delta R_s^y - \Delta R_s^x)/2$ can be obtained by comparing the highest values of the two curves in Figure 1(c).

For the measurement in Figure 1(d), as the sample stage is rotated inside a magnetic field, the NLSV remains in the same measurement circuit. A clear periodic signal change is observed and the apparent value for $(\Delta R_s^y - \Delta R_s^x)/2$ is 0.17 mΩ, with the $R_s$ of 90° ($y$-spins) being higher than that of 0° ($x$-spins). However, the rotation of the sample stage may cause subtle changes of wiring configuration in the circuit thereby changing the $R_s$ baseline value. To compensate for this, the rotational measurement is carried out in zero magnetic field, as shown in Figure S1, and a small periodic signal of – 0.03 mΩ is observed. The negative sign indicates that the $R_s$ value of 0° is higher than that of 90°, opposite to the situation in Figure 1(d). This observed periodic signal must come from the baseline change associated with the rotation, because the spins in the NLSV do not change relative to the sample coordinate as the sample rotates in the zero magnetic field. Therefore, this signal of – 0.03 mΩ should be subtracted from the apparent signal of 0.17 mΩ in Figure 1(d), and the adjusted anisotropic signal is $(\Delta R_s^y - \Delta R_s^x)/2$ = 0.17 mΩ - (-0.03 mΩ) = 0.20 mΩ.

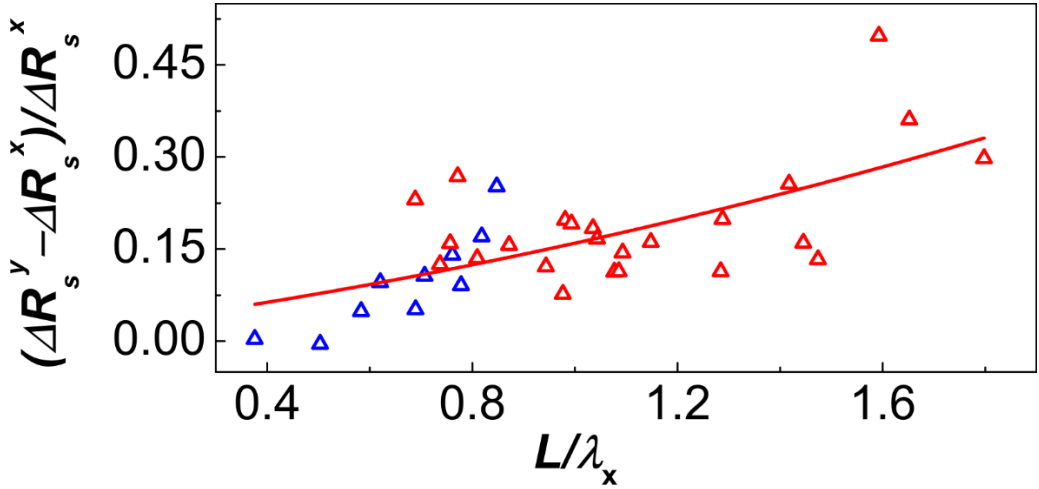

**Figure S2.** The plot of $(\Delta R_s^y - \Delta R_s^x)/\Delta R_s^x$ versus $L/\lambda_x$ for NLSVs on both substrates A (blue symbols) and B (red symbols). The solid line is guidance to the eyes.

The measurements of $(\Delta R_s^y - \Delta R_s^x)/2$ in Figure 2 is carried out in a probe station. The electrical contacts to the NLSV and the entire detection circuit remain intact during the measurement, as the net magnetic field alternates between $x$ and $y$ directions. Therefore, the baseline should indeed remain unchanged during the measurement and the detected $(\Delta R_s^y - \Delta R_s^x)/2$ in Figure 2(d) is accurate.

**Notes S2. Combined plot of $(\Delta R_s^y - \Delta R_s^x)/\Delta R_s^x$ versus $L/\lambda_x$ for substrates A and B.**

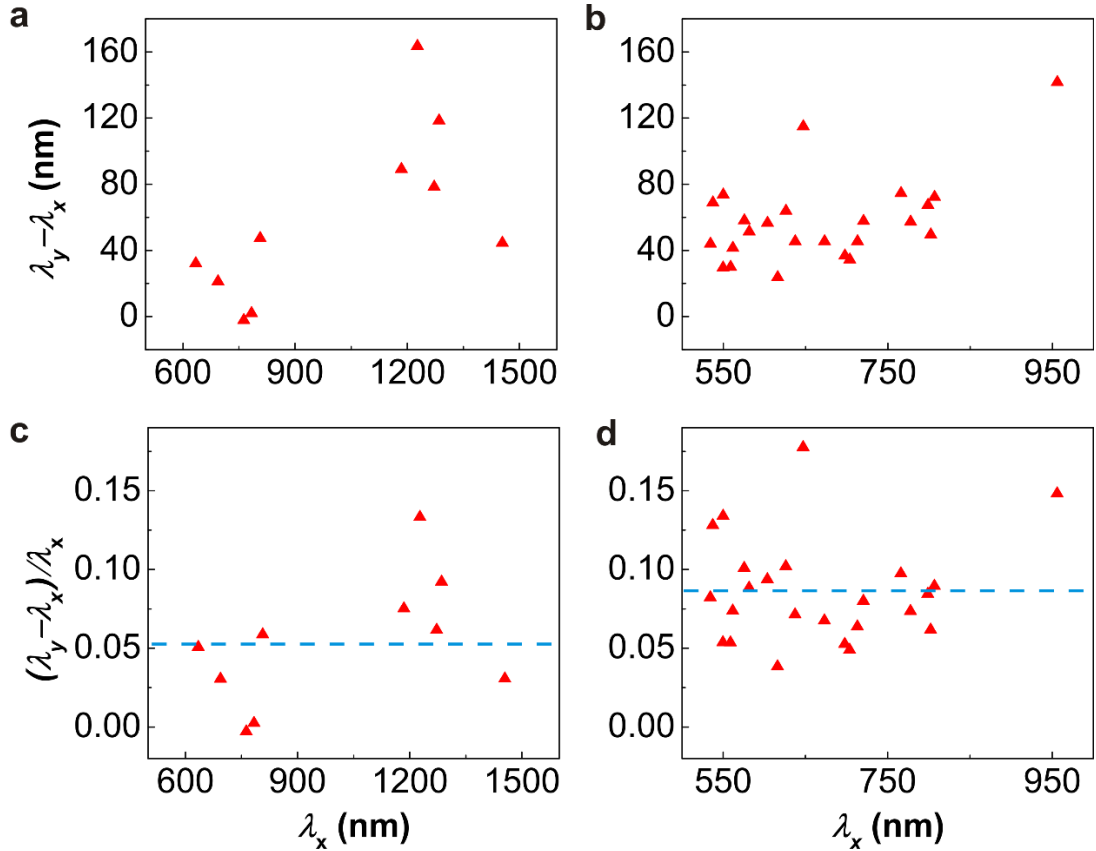

**Figure S3.** The estimated anisotropic change in spin diffusion length $(\lambda_y - \lambda_x)$ at 10 K is plotted as a function of estimated $\lambda_x$ for NLSVs on (a) substrate A and (b) substrate B. The percentage change $(\lambda_y - \lambda_x)/\lambda_x$ versus $\lambda_x$ at 10 K for (c) substrate A and (d) for substrate B. The dashed lines indicate average values.

Figures 3(d) and 3(f) in the main article show the percentage anisotropic signal $(\Delta R_s^y - \Delta R_s^x)/\Delta R_s^x$ versus normalized channel length $L/\lambda_x$ plots for NLSVs on substrates A and B, respectively. The data in these two plots are now combined to generate a composite plot in Figure S2 for all NLSVs on these two substrates. There is a clear systematic increase of $(\Delta R_s^y - \Delta R_s^x)/\Delta R_s^x$ as a function of $L/\lambda_x$, which is qualitatively consistent with the calculation shown in Figure 3(b). This affirms that the anisotropic signal originates from the anisotropic spin-relaxation in the Cu channels instead of the anisotropic spin polarization of the $F_1$ and $F_2$.

The average values of $(\Delta R_s^y - \Delta R_s^x)/\Delta R_s^x$ are lower for substrate A than for substrate B, because the average $\lambda_x$ is longer for substrate A and the average normalized channel length $L/\lambda_x$ is lower.

**Notes S3. Estimated anisotropic change of spin-relaxation lengths**

We have provided evidence in the main article that that anisotropic signal is a direct result of anisotropic difference between the Cu spin-relaxation lengths $\lambda_x$ for x-spins and the $\lambda_y$ for y-spins. For each NLSV, the values of $\lambda_y$ can be estimated from the spin signals of y-spin $\Delta R_s^y$ in a way similar to how $\lambda_x$ is obtained from $\Delta R_s^x$ in the main article. More specifically we assume $P = 20\%$ and calculate $\lambda_y$ from Eq. (1) using measured $\rho$, $L$, and $A$ for that NLSV. The obtained anisotropic difference of spin-relaxation length $(\lambda_y - \lambda_x)$ is plotted as a function of the $\lambda_x$ for NLSVs on substrates A and B, as shown in Figure S3(a) and S3(b), respectively. The percentage change $(\lambda_y - \lambda_x)/\lambda_x$ is shown in Figure S3(c) and S3(d) as a function of $\lambda_x$ for substrates A and B, respectively. The average value is $(\lambda_y - \lambda_x)/\lambda_x = (5.3 \pm 4.1)\%$ for substrate A and $(\lambda_y - \lambda_x)/\lambda_x = (8.6 \pm 3.3)\%$ for substrate B.

**Notes S4. Computational details of spin injection into bulk Cu.**

We construct a bilayer along face-centered cubic (111) crystal plane consisting of 10-nm-thick Py and 100-nm-thick Cu, both of which are made diffusive by imposing the frozen thermal lattice disorder.[55] The disorder is implemented in a 5×5 supercell in the lateral directions with periodic boundary condition. The electronic structure of the bilayer is calculated self-consistently based upon the local density approximation of density functional theory. The lattice constant of Cu, namely 3.614 Å, is used for the whole system. The potentials, charge and spin densities of Fe and Ni atomic spheres in Py are obtained within the coherent potential approximation[56] implemented with the minimal basis of tight-binding linear muffin-tin orbitals[57]. The magnetic moments in the Fe and Ni atomic spheres are 2.71 and 0.64 Bohr magneton, respectively.

Using semi-infinite ballistic Cu electrodes in both sides of the bilayer, the transport calculation is performed with the Pauli spin-orbit Hamiltonian perturbatively included. The strength of the spin-orbit interaction is determined from the potential gradient within the atomic spheres. For the calculation with artificially increased spin-orbit strength, we merely multiply the real spin-orbit strength by a factor of 7 for Cu atomic spheres in the scattering region[58]. The scattering states at the Fermi level are obtained using the so-called "wave-function matching" formalism[59] and the position-dependent spin accumulation is determined by calculating the local spin polarization of electrical current density[60].

**Notes S5. Qualitative description of spin-relaxation rate due to surface spin-orbit effects.**

We take the Cartesian coordinate as shown in Figure 4(b) in the main article and the spins diffuse along the $y$-axis in the Cu channel. In the bulk region, where conduction electrons are not influenced by surface scattering, spin-relaxation rate is isotropic with respect to the spin orientation. When conduction electrons are close to a surface of the Cu channel, spin-orbit (SO) interaction at the surface results in an additional spin-relaxation mechanism.

Here we consider the spin-relaxation rates of $x$ and $y$-spins under the perturbation of the Rashba-Sheka-Vasko type[21,22] SO Hamiltonians at the top/bottom and side surfaces, respectively. The SO Hamiltonian of the top/bottom surfaces lying in the $x$-$y$ plane can be expressed as

$$H_R^{t/b} = \frac{\alpha_R}{\hbar} \boldsymbol{\sigma} \cdot (\hat{z} \times \boldsymbol{p}) = \frac{\alpha_R}{\hbar}\left(p_x \sigma_y - p_y \sigma_x\right) = \frac{\alpha_R}{\hbar}\begin{pmatrix} 0 & -ip_x - p_y \\ ip_x - p_y & 0 \end{pmatrix}, \quad (S1)$$

where $\alpha_R$ is the strength of surface SO interaction, $\boldsymbol{\sigma}$ is Pauli matrices and $\boldsymbol{p}$ is the momentum of electrons. The $x$-spins are represented as the eigenstates of $s_x$, i.e. $|\pm x\rangle = (1, \pm 1)/\sqrt{2}$. The spin-relaxation rate (inversely proportional to the spin-relaxation time) of the $x$-spins due to the SO Hamiltonian Eq. (S1) can be calculated via the Fermi golden rule,[61]

$$\frac{1}{\tau_x^{t/b}} \propto \int d^3 p \left| \langle +x | H_R^{t/b} | -x \rangle \right|^2 = \frac{\alpha_R^2}{\hbar^2} \int d^3 p \cdot p_x^2. \tag{S2}$$

The spin-relaxation rate of the $y$-spins $|\pm y\rangle = (1, \pm i)/\sqrt{2}$ can be obtained in the same manner,

$$\frac{1}{\tau_y^{t/b}} \propto \int d^3 p \left| \langle +y | H_R^{t/b} | -y \rangle \right|^2 = \frac{\alpha_R^2}{\hbar^2} \int d^3 p \cdot p_x^2. \tag{S3}$$

Here we find $1/t_x^{t/b} = 1/t_y^{t/b}$ indicating that the SO effect at the top/bottom surfaces does not introduce anisotropic spin-relaxation rate for $x$- and $y$-spins.

The side surfaces of the Cu channel offer another source of SO interaction,

$$H_R^s = \frac{\alpha_R}{\hbar} \boldsymbol{\sigma} \cdot (\hat{x} \times \boldsymbol{p}) = \frac{\alpha_R}{\hbar} (p_y \sigma_z - p_z \sigma_y) = \frac{\alpha_R}{\hbar} \begin{pmatrix} p_y & ip_z \\ -ip_z & -p_y \end{pmatrix}. \tag{S4}$$

The spin-relaxation rates of $x$- and $y$-spins due to Eq. (S4) are given by

$$\frac{1}{\tau_x^s} \propto \int d^3 p \left| \langle +x | H_R^s | -x \rangle \right|^2 = \frac{\alpha_R^2}{\hbar^2} \int d^3 p \cdot (p_y^2 + p_z^2), \tag{S5}$$

$$\frac{1}{\tau_y^s} \propto \int d^3 p \left| \langle +y | H_R^s | -y \rangle \right|^2 = \frac{\alpha_R^2}{\hbar^2} \int d^3 p \cdot p_y^2. \tag{S6}$$

Therefore, the *y*-spins have a smaller spin-relaxation rate and hence a longer spin-relaxation length than the *x*-spins. The Rashba-Sheka-Vasko SO effect at the side surfaces of the Cu channel indeed results in anisotropic spin-relaxation in agreement with our experimental measurement.